\documentclass[10pt,aps,showpacs,unsortedaddress,superscriptaddress, twocolumn]{revtex4-2}

\usepackage{graphicx}
\usepackage{dcolumn}
\usepackage{bm}
\usepackage{amsmath}
\usepackage{amsfonts}
\usepackage{amssymb}%
\usepackage{mathtools}
\usepackage{braket}
\usepackage{comment}
\usepackage{multirow}

\usepackage{here} %

\usepackage{bm}

\usepackage{color}

\usepackage{xcolor}
\DeclareRobustCommand{\erase}{\bgroup\markoverwith{\textcolor{black}{\rule[.5ex]{2pt}{0.4pt}}}\ULon}

\newcommand{\1}{\mathrm{I}}
\newcommand{\2}{\mathrm{I}\hspace{-1.2pt}\mathrm{I}}
\newcommand{\3}{\mathrm{I}\hspace{-1.2pt}\mathrm{I}\hspace{-1.2pt}\mathrm{I}}
\newcommand{\4}{\mathrm{I}\hspace{-1.2pt}\mathrm{V}}
\newcommand{\5}{\mathrm{V}}

\usepackage{siunitx}

\usepackage{amsthm}

\theoremstyle{plain} 
\newtheorem{thm}{Theorem}
\newtheorem{lem}[thm]{Lemma}
\newtheorem{prop}{Proposition}

\providecommand{\U}[1]{\protect\rule{.1in}{.1in}}

\begin{document}

\preprint{APS/123-QED}%

\title{Finite-size security proof of binary-modulation continuous-variable 
\\quantum key distribution using only heterodyne measurement}

\author{Shinichiro Yamano}%
\email{yamano@qi.t.u-tokyo.ac.jp}
\affiliation{
Department of Applied Physics, Graduate School of Engineering,The University of Tokyo, 7-3-1 Hongo Bunkyo-ku, Tokyo 113-8656, Japan}

\author{Takaya Matsuura}
\affiliation{
Department of Applied Physics, Graduate School of Engineering,The University of Tokyo, 7-3-1 Hongo Bunkyo-ku, Tokyo 113-8656, Japan}

\author{\\Yui Kuramochi}
\affiliation{
Department of Physics, Kyushu University, 744 Motooka, Nishi-ku, Fukuoka, Japan}

\author{Toshihiko Sasaki}
\affiliation{
Department of Applied Physics, Graduate School of Engineering,The University of Tokyo, 7-3-1 Hongo Bunkyo-ku, Tokyo 113-8656, Japan}
\affiliation{
Photon Science Center, Graduate School of Engineering,The University of Tokyo, 7-3-1 Hongo, Bunkyo-ku, Tokyo 113-8656, Japan}

\author{Masato Koashi}
\affiliation{
Department of Applied Physics, Graduate School of Engineering,The University of Tokyo, 7-3-1 Hongo Bunkyo-ku, Tokyo 113-8656, Japan}
\affiliation{
Photon Science Center, Graduate School of Engineering,The University of Tokyo, 7-3-1 Hongo, Bunkyo-ku, Tokyo 113-8656, Japan}

\begin{abstract}%
Continuous-variable quantum key distribution (CV-QKD) has many practical advantages including compatibility with current optical communication technology. Implementation using heterodyne measurements is particularly attractive since it eliminates the need for active phase locking of the remote pair of local oscillators, but the full security of CV QKD with discrete modulation was only proved for a protocol using homodyne measurements. Here we propose an all-heterodyne CV-QKD protocol with binary modulation and prove its security against general attacks in the finite-key regime. Although replacing a homodyne measurement with a heterodyne measurement would be naively expected to incur a 3-dB penalty in the rate-distance curve, our proof achieves a key rate with only a 1-dB penalty.
\end{abstract}
\maketitle

\section{\label{sec:level1}Introduction \lowercase{}} 
Quantum key distribution (QKD) is the technology that enables information-theoretically secure communication between two separate parties.
QKD is classified into two categories: discrete-variable (DV) QKD and continuous-variable (CV) QKD. 
DV-QKD protocols often encode information to a photon in different optical modes
such as different polarizations or time bins. They use photon detectors to read out the encoded information.
This type has a long history since early studies \cite{BENNETT1984,Bennett1992}.
A lot of knowledge about the finite-key analysis and how to handle imperfections of actual devices has been accumulated.
On the other hand, CV-QKD protocols encode information to quadrature in the phase space of an optical pulse.
They use homodyne or heterodyne detection \cite{Ralph1999,Grosshans2002},
which is highly compatible with the coherent optical communication technology currently widespread in industry \cite{eriksson2020wavelength, huang2015continuous, kumar2015coexistence, huang2016field, karinou2017experimental, karinou2018toward, eriksson2018coexistence, eriksson2019wavelength}.
See Refs.~\cite{xu2020secure,pirandola2020advances} for comprehensive reviews of the topic.

To implement the homodyne measurement, the local oscillator (LO) phases of the sender (Alice) and the receiver (Bob) should be matched. Since they are independently drifted in the actual experiment, they have to be calibrated continuously.
For this purpose, the so-called local LO scheme is mainly used \cite{Qi2015,Soh2015} in the implementation. 
In this scheme, using pilot pulses as a phase reference, the relative phase between the two remote LOs is tracked and corrected.
This real-time feedback scheme, however, complicates Bob's receiver and makes it difficult to be integrated into the conventional systems \cite{Diamanti2015,Huang:15,Marie2017,Wang:18,Wonfor2019}.

On the other hand, in the heterodyne measurement, the difficulty related to the phase locking is greatly reduced. The heterodyne measurement outputs two quadrature amplitudes, which contain the information about the relative phase to the LO. As shown in some experiments \cite{https://doi.org/10.48550/arxiv.1908.03625,Huang:20}, Bob can compensate for the phase mismatch between the two LO after the measurement. Therefore the CV-QKD protocol using only heterodyne measurement removes the need for the real-time phase-compensation system and makes
the implementation easier.

In terms of security analysis, due to difficulty in dealing with continuous observables,
the security of CV-QKD was proved only under limited conditions, such as specific attacks \cite{papanastasiou2017finite, papanastasiou2021continuous, samsonov2020subcarrier} and asymptotic cases \cite{leverrier2009unconditional, zhao2009asymptotic, bradler2018security, lin2019asymptotic, ghorai2019asymptotic, namiki2022security}.
The full security in the finite-size regime against general attacks was then proved for a Gaussian modulation protocol \cite{leverrier2013security, leverrier2017security}, but it did not cover discretization of modulation necessary in actual implementation \cite{jouguet2012analysis, lupo2020towards, kaur2021asymptotic}. 
Recently, a binary phase modulated CV-QKD protocol with a full security proof was reported \cite{m}.
It uses both homodyne and heterodyne measurements which should be actively switched. In this protocol, a homodyne measurement is used for generating a raw key and a heterodyne measurement for monitoring the attacks. Since this protocol involves homodyne measurements, it has the phase locking problem mentioned above.

In this paper, we propose a finite-key analysis of a CV-QKD protocol with binary phase modulation that uses only heterodyne measurements. 
The heterodyne measurement consists of two homodyne measurements whose inputs are made by splitting the original input into two halves.
Due to this apparent $3$-dB loss, straightforward application of the security proof
in Ref.~\cite{m} to the all-heterodyne protocol may suffer from a $3$-dB penalty in the key rate as a function of distance. Our security proof here shows that
the penalty in the key rate can be suppressed to be only about $\SI{1}{\decibel}$.
Moreover, we show that the security can be guaranteed even if we simplify the protocol by omitting the random discarding of rounds required in Ref.~\cite{m}.

The article is organized as follows.
In Sec.~$\2$, we introduce our protocol with only heterodyne measurements.
In Sec.~$\3$, we provide a sketch of the security proof based on analyzing the statistics of phase errors in a virtual protocol, while we describe the detail of the proof in Methods.
Numerical simulations of the key rates as a function of distance are given in Sec.~$\4$.
In Sec.~$\5$, we discuss how our proof mitigates the apparent $3$-dB penalty in the key rate.

\section{\label{sec}Protocol \lowercase{}}

\subsection{\label{sec:level2}Proposed Protocol} 
\begin{figure}[htbp]
\centering
\includegraphics[width=9cm,angle=0,clip]{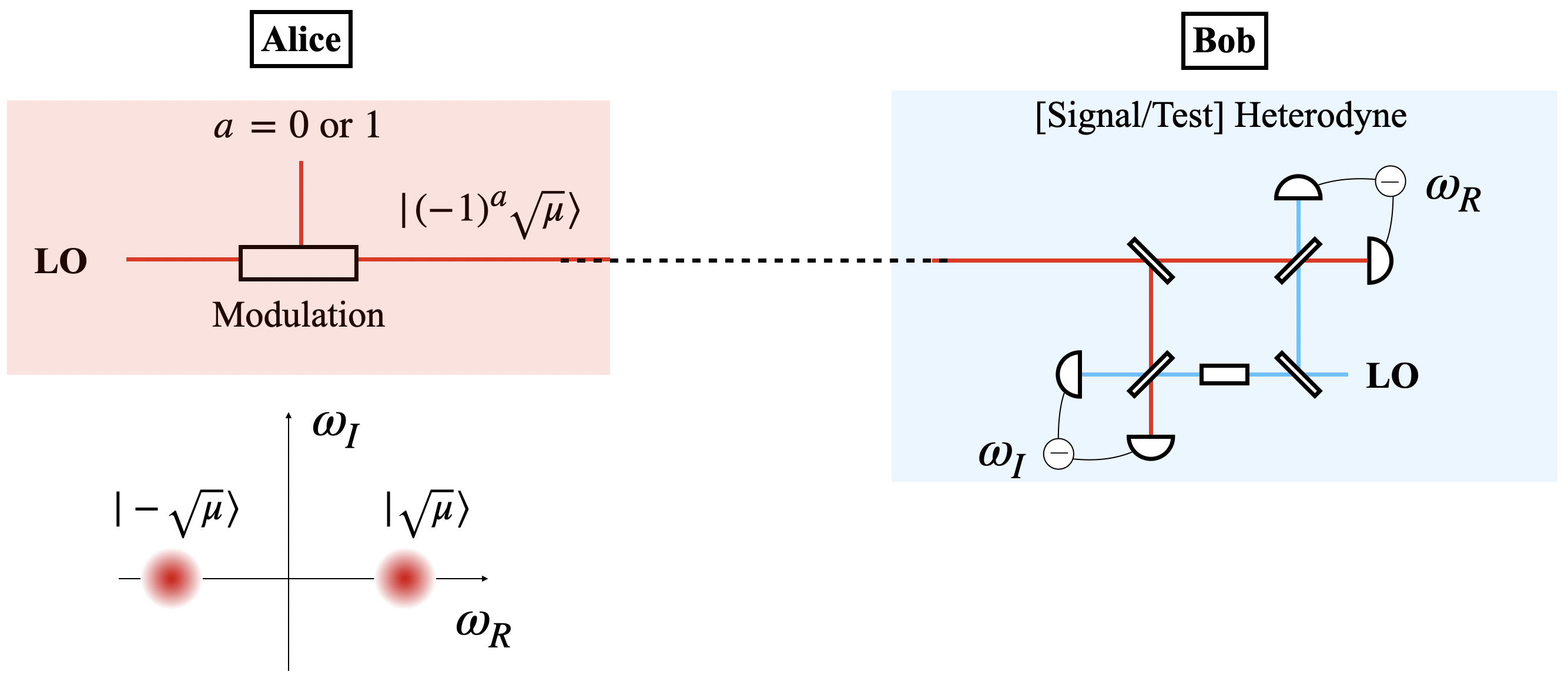}
\caption{Alice sends coherent state $\ket{\sqrt{\mu}}$ or $\ket{-\sqrt{\mu}} $ randomly. Bob performs a heterodyne measurement. After $N$ rounds of communication, Bob randomly decides whether the role of each round is ``signal'' or ``test''.
Alice and Bob use independent LOs (local oscillators).
Bob's outcome $(\hat{\omega}_R,\hat{\omega}_I)$ can be compensated when Alice and Bob learns the phase difference between their LOs later.}
\label{fig0}
\end{figure}
We describe our protocol as follows (see Fig.~\ref{fig0}).
In the description, the outcome of the heterodyne measurement is represented by a complex number that is normalized such that its mean coincides with the complex mean amplitude of the input. The definition of the function $\Lambda_{m,r}$ will be given in the next subsection. The binary entropy function is defined by $h(x)\coloneqq -x\log_2(x)-(1-x)\log_2(1-x)$.
\\
\\
{\bf Actual Protocol}

Alice and Bob predetermine the protocol parameters [$N, \epsilon, \mu , p_{\rm sig} , p_{\rm test} , \beta , s , s' , \kappa , \gamma, m, r$]
and acceptance functions $f_{\rm suc,0}(x) $ and $f_{\rm suc,1}(x)$ which map the real number $\mathbb{R}$ into the closed interval $[0,1]$.
Here $N,s,s'$ are positive integers, $m$ is a positive odd integer, $\mu,r,\beta,\epsilon > 0,$ $\kappa,\gamma \geq 0$,  $p_{\rm sig}, p_{\rm test}\in [0,1]$,  $p_{\rm sig}+p_{\rm test}=1$,  $f_{\rm suc,1}(x)=f_{\rm suc,0}(-x)$, and $f_{\rm suc,1}(x) + f_{\rm suc,0}(x) \le 1$.
\\
\\
1.\ Alice randomly chooses a bit $a \in  \{0 ,1 \}$.
 She sends an optical pulse $\tilde C$ in the coherent state with complex amplitude $(-1)^a\sqrt{\mu}$ to Bob.
 She repeats it $N$ times.
\\
\\
2.\ On each pulse $C$ of the received $N$ pulses, Bob performs a heterodyne measurement and obtains an outcome $\hat\omega = \hat\omega_R +i \hat\omega_I\ ( \hat\omega_R, \hat\omega_I \in \mathbb{R})$. 
\\
\\
3.\ 
Alice and Bob processes the raw data associated with each of the $N$ transmissions, which we call a round, in the following way. Bob randomly chooses the role of the round and announces it such that a ``signal round'' is chosen with probability $p_{\rm sig}$ and a ``test round'' with $p_{\rm test}$.
According to the announced role, Alice and Bob do one of the following procedures.
\\
\\
\lbrack\,signal\,\rbrack\, 
Bob determines the bit $b\in \{0 ,1 \}$ or ``failure'' according to the real part $\hat\omega_R$ of the measurement outcome, such that the probability for each event is given by the acceptance functions as follows:
\begin{align}
\label{pr1}
{\rm Pr}(b=0)&= f_{\rm suc,0}(\hat\omega_R),\\
\label{pr2}
{\rm Pr}(b=1)&= f_{\rm suc,1}(\hat\omega_R) = f_{\rm suc,0}(-\hat\omega_R),\\
\label{pr3}
{\rm Pr}({\rm failure})&= 1- f_{\rm suc,0}(\hat\omega_R) - f_{\rm suc,1}(\hat\omega_R).
\end{align}
Bob announces  ``success'' when he obtained the bit $b \in \{ 0,1\}$ and announces ``failure'' otherwise.
If he announces ``failure'', Alice discards her bit $a$.
\\
\\
\lbrack\,test\,\rbrack\,Alice announces the bit $a$ to Bob and he calculates the value  $\Lambda_{m,r}(|\hat\omega - (-1)^a\beta|)$.
\\
\\
4.\ The $N$ rounds are divided into ``signal-success'', ``signal-failure'' and ``test'' rounds, whose numbers
are denoted by $\hat N^{\rm suc}$,$\hat N^{\rm fail}$ and $\hat N^{\rm test}$, respectively. The number of ``signal'' rounds is denoted by $\hat N^{\rm sig} \coloneqq \hat N^{\rm suc}  + \hat N^{\rm fail}$. Alice and Bob concatenate their own bits kept in the signal-success rounds to define $\hat N^{\rm suc}$-bit sifted keys. Bob calculates the sum of $\Lambda_{m,r}(|\hat\omega - (-1)^a\beta|)$ in all test rounds, which is denoted by $\hat F$.
\\
\\
5.\ Alice and Bob perform bit error correction on the sifted keys. They consume $H_{\rm EC}$ bits of the pre-shared key for privately transmitting the syndrome of a linear code and $s'$ bits of that for the verification.
\\
\\
6.\ 
Alice and Bob perform privacy amplification on the $\hat N^{\rm suc}$-bit reconciled keys.
The length is shortened by $\hat N^{\rm suc}h(U(\hat F)/\hat N^{\rm suc})+s$, where the function $U(\hat F)$ will be given in Eq.~\eqref{form-of-U}.
The final key length $\hat N^{\rm fin}$ is thus given by
\begin{align}
 \label{final-key-length}
 \hat N^{\rm fin} = \hat N^{\rm suc}\left(1 - h(U(\hat F)/\hat N^{\rm suc})\right) -s.
\end{align}

In the above protocol, the net key gain per pulse is given by
\begin{align}
\hat{G} =  \left( \hat N^{\rm fin} - H_{\rm EC} -s' \right)/N.
\end{align}
The above protocol is feasible even when Alice's and Bob's LOs are not phase-locked, 
as long as one can provide a good guess on their relative phase for each pulse.
In such a case, the outcome $\hat{\omega}$ in Step 2 is determined as follows.
First, Bob records the complex amplitude $\hat{\omega}'$ obtained 
directly from his heterodyne measurement. Then, using the guess on the relative phase, 
he appropriately choose an angle $\theta$ to define a compensated value 
$\hat{\omega}\coloneqq e^{i\theta}\hat{\omega}'$ such that the 
real axis of $\hat{\omega}$ coincides with the direction of Alice's binary modulation. 
\subsection{Definition of $\Lambda_{m,r}$ and its property}
The function $\Lambda_{m,r}$ relates the outcome of the
heterodyne measurement to a bound on the fidelity of the input state to a coherent state \cite{m}, see also Ref.~\cite{chabaud2021efficient}.
It is defined as
\begin{align}
 \Lambda_{m,r}(\mu) \coloneqq  e^{-r\mu}(1+r)L^{(1)}_{m}((1+r)\mu),
\end{align}
where $L^{(1)}_{m}$ is the associated Laguerre polynomial
\begin{align}
L_{n}^{(k)}(\nu) \coloneqq (-1)^k\frac{d^k L_{n+k}(\nu)}{d\nu^k},
\end{align} 
and $L_{n}(\nu)$ is the Laguerre polynomial
\begin{align}
 L_{n}(\nu) \coloneqq \frac{e^\nu}{n!}\frac{d^n}{d\nu^n}(e^{-\nu}\nu^n).
\end{align}

For a state $\rho$ of an optical pulse, the heterodyne measurement produces an outcome $\hat{\omega} \in \mathbb{C}$ with a probability measure
\begin{align}
\label{rho_dense}
q_{\rho}(\omega)d^2\omega \coloneqq \bra{\omega}\rho\ket{\omega}\frac{d^2\omega}{\pi},
\end{align}
where $\ket{\omega}$ is a coherent state
\begin{align}
\ket{\omega} \coloneqq e^{-|\omega|^2/2}\sum_{n = 0}^{\infty} \frac{\omega^n}{\sqrt{n!}}\ket{n}.
\end{align} 
The expectation value of a function $f(\omega)$ based on the probability measure in Eq.~(\ref{rho_dense}) is denoted as $\mathbb{E}_{\rho}[f(\hat{\omega})]$.
In Ref.~\cite{m}, it has been shown that a fidelity $\bra{\beta}\rho\ket{\beta}$ satisfies the following relation
\begin{align}
\label{lag}
 \mathbb{E}_{\rho}[\Lambda_{m,r}(|\hat{\omega}- \beta|^2)] \leq{\rm Tr}(\rho \ket{\beta}\bra{\beta}),\hspace{15pt} (m : {\rm odd} ).
\end{align}

\section{Sketch of the security proof}
The finite-size security of the proposed protocol against general attacks
can be shown using
the Shor and Preskill approach \cite{Shor2000}, similar to Ref.~\cite{m}.
This approach connects the amount of the privacy amplification
to the so-called phase error rate.
Here, we describe the sketch of the security proof.
We mainly focus on the intuitive explanation on how we can bound the number of the phase errors and
we also comment on differences between our proof and
that of Ref.~\cite{m}.
The full proof will be described in Methods.

First, we define the phase error.
For that purpose,
we introduce an entanglement-sharing protocol
in which
Alice and Bob share $\hat N^{\rm suc}$ pairs of qubits in the signal-success rounds in such a way that measuring those qubits should produce binary sequences equivalent to the sifted keys in Actual protocol.   
A major difference from Ref.~\cite{m} appears in how we define Bob's qubits, because Bob's sifted key bit is generated from a heterodyne measurement in our protocol instead of a homodyne measurement in Ref.~\cite{m}.
\\
\\
{\bf Entanglement-sharing protocol}
\\
1.\ Alice prepares a qubit $A$ and an optical pulse $\tilde{C}$  in the state $\ket{\Psi}_{A\tilde C} $ defined as
\begin{align}
\label{psiA}
\ket{\Psi}_{A\tilde C}  \coloneqq \frac{\ket{0}_{A}\ket{\sqrt{\mu}}_{ \tilde{C} }+\ket{1}_{A}\ket{-\sqrt{\mu}}_{\tilde{C}}}{\sqrt{2}},
\end{align}
and sends $\tilde C$ to Bob. She repeats it $N$ times.
\\
\\
2.\ For each of the $N$ rounds, with the probabilities $p_{\rm sig}$ and $p_{\rm test}$, Bob determines whether each round is ``signal'' or ``test'' and announces it.
Based on this label, Alice and Bob proceed as follows.
\\
\\
\lbrack\,signal\,\rbrack\,Bob performs a quantum operation (specified by trace-non-increasing and completely positive map) $\mathcal{F}$ on the received optical pulse $C$, where
\begin{align}
\label{F()}
 \mathcal{F}(\rho) &\coloneqq \int_{\mathbb{C}} d^2\omega K(\omega)\rho K^{\dagger}(\omega), 
 \end{align}
 \begin{align}
 K(\omega) &\coloneqq \sqrt{\frac{f_{{\rm suc},0}(\omega_R)}{\pi}}  \ket{0}_{B}\bra{\omega}_{C} + \sqrt{\frac{f_{{\rm suc},1}(-\omega_R)}{\pi}}\ket{1}_{B}\bra{-\omega}_{C} \\
 \label{actCP}
  &= \sqrt{\frac{f_{{\rm suc},0}(\omega_R)}{\pi}}(\ket{0}_{B}\bra{\omega}_{C} +\ket{1}_{B}\bra{-\omega}_{C}  ).
\end{align}
This operation heralds ``success'' or ``failure'', which Bob announces, and in the former case produces a qubit $B$ in the state $\mathcal{F}(\rho)/{\rm Tr}(\mathcal{F}(\rho))$. 
\\
\\
\lbrack\,test\,\rbrack\,Bob performs a heterodyne measurement and obtains an outcome $\hat\omega = \hat\omega_R +i \hat\omega_I\ ( \hat\omega_R, \hat\omega_I \in \mathbb{R})$. Alice measures her qubit $A$ on the $Z$ basis (\{$\ket{0}, \ket{1}$\}) and announces the outcome $a\in\{ 0,1 \}$ to Bob, who calculates $\Lambda_{m,r}(|\hat\omega - (-1)^a\beta|)$ as in Actual protocol.
\\
\\
3.\ Alice and Bob define $\hat N^{\rm suc}$,$\hat N^{\rm fail}, \hat N^{\rm test}, \hat N^{\rm sig}$, and $\hat F$ as in Actual protocol. At this point, Alice and Bob share $\hat N^{\rm suc}$ qubits. 
\\

This entanglement-sharing protocol can be made equivalent to Actual protocol 
by measuring the $\hat N^{\rm suc}$ pairs of qubits left by the protocol on the $Z$ basis. This can be confirmed by the following two observations. First, measuring
the qubit $A$ of the state $\ket{\Psi}_{A\tilde C}$ on the $Z$ basis reproduces the bit $a$ as well as the state of the optical pulse $\tilde{C}$ of Actual protocol.
Second, Eqs.~(\ref{F()})-(\ref{actCP}) lead to
\begin{align}
 \bra{b}  \mathcal{F}(\rho) \ket{b} =&  \int_{\mathbb{C}}d^2\omega \; f_{{\rm suc},b}(\omega_R)\frac{\bra{\omega}\rho_C\ket{\omega}}{\pi},
\end{align}
which shows that Bob's procedure of determining the bit $b$ is equivalent to that in Actual protocol.

Based on this entanglement-sharing protocol, we define the phase error as follows.
After Entanglement-sharing protocol, suppose that
Alice and Bob measure each of their $\hat N^{\rm suc}$ qubits on the $X$ basis \{$\ket{\pm} \coloneqq \frac{\ket{0}\pm\ket{1}}{\sqrt{2}} $\}
instead of the $Z$ basis.
Each outcome can be denoted by $+$ or $-$.
The pair of Alice's and Bob's outcomes can thus be written as an element in $\{(x_A,x_B)| x_A,x_B\in\{+,-\}\}$.
We call the outcomes ($+,-$) and ($-,+$) as a ``phase error''.
The number of phase errors is denoted by $\hat N_{\rm ph}$.

It is known \cite{Hayashi2012,Koashi2009} 
that if we can find a function $U$
of the data observed in the test rounds that bounds $\hat{N}_{\rm ph}$ from above,
we can derive a sufficient amount of the privacy amplification to
achieve the required secrecy of the protocol. In our case, if we can
find $U$ that satisfies
\begin{align}
\label{phase_ineq}
{\rm Pr}\left(\hat{N}_{\rm ph} \le U(\hat{F})     \right) \ge 1-\epsilon 
\end{align}
in Entanglement-sharing protocol followed by $X$-basis
measurements on the $\hat{N}^{\rm suc}$ qubits, our protocol is
$(\sqrt{2}\sqrt{\epsilon+2^{-s}}+2^{-s'})$-secure.

For the construction of $U(\hat F)$, we regard 
occurrence of a phase error as an outcome of a generalized measurement on Alice's qubit $A$
and the optical pulse $C$ received by Bob.
The positive-operator-valued measure (POVM) for this measurement is constructed as follows. Bob's measurement has three outcomes, 
$(+, -, {\rm failure})$, whose POVM elements are denoted, respectively, by 
$(M_{\rm ev}, M_{\rm od}, M_{\rm fail})$.  The use of ev/od is because the outcomes $+$ and $-$ implies even and odd photon numbers, respectively (see Methods).  The explicit form of the operators can be determined from the relations 
${\rm Tr}(M_{\rm ev} \rho_C)=\bra{+}\mathcal{F} (\rho_C) \ket{+}$,
${\rm Tr}(M_{\rm od} \rho_C)=\bra{-}\mathcal{F} (\rho_C)\ket{-}$ and 
$M_{\rm fail}={\bm 1}_C - M_{\rm ev} - M_{\rm od}$. 
The POVM elements $M_{x_A, x_B}$ of the outcome $(x_A, x_B)$ for Alice's and Bob's $X$-basis measurement is then 
given by 
\begin{align}
\label{aaa1}
  M_{x_A, +(-)}=\ket{x_A}\bra{x_A}\otimes M_{\mathrm{ev(od)}}.
\end{align}
The phase error is then represented by the operator 
\begin{align}
\label{aaa2}
  M_{\rm ph} = M_{+,-}+M_{-,+}.
\end{align}

As an intuitive explanation, let us consider an asymptotic limit of $N \to \infty$, 
in which Eve's attack is fully characterized by the state $\rho_{{\rm out}, AC}$ for Alice's qubit $A$
and the optical pulse $C$ received by Bob, averaged over the $N$ rounds.
In this limit, $\hat N_{\rm ph}/N p_{\rm sig}$ converges to ${\rm Tr}(M_{\rm ph} \rho_{{\rm out}, AC})$
and hence finding an upper bound $U(\hat F)$ in Eq.~(\ref{phase_ineq})
amounts to finding an upper bound on ${\rm Tr}(M_{\rm ph} \rho_{{\rm out}, AC})$.

The state $\rho_{{\rm out}, AC}$ is restricted by two conditions about the
input states and about the observed data in the test rounds.
The first condition comes from the fact that the reduced state of Alice's qubit $A$
is unaffected by Eve's attack, implying that ${\rm Tr}_C \rho_{{\rm out}, AC}$
is the same as the reduced state of the initial state $\ket{\Psi}_{A\tilde C} $
in Eq.~(\ref{psiA}). It leads to a constraint written as 

\begin{align}
\label{defq-}
 {\rm Tr}_{AC}\left(\rho_{\mathrm{out},AC} \Pi_{-}^{\rm\textcolor{black}{sig}} \right) =q_-,
 \end{align}
 where
\begin{align}
\label{Pi_sig}
\Pi_{-}^{\rm\textcolor{black}{sig}}\coloneqq\ket{-}\bra{-}_A\otimes\mathbf{1}_C
\end{align}
and
 \begin{align}
 \label{Eq_q}
q_-\coloneqq & {\rm Tr} \left( \ket{\Psi}\bra{\Psi}_{A\tilde{C}} \Pi_{-}^{\rm\textcolor{black}{sig}} \right) \nonumber \\
=& \frac{1}{2}\left(1-\braket{\sqrt{\mu}\mid-\sqrt{\mu}}\right) \nonumber \\
=& \frac{1}{2}\left( 1-e^{-2\mu} \right).
\end{align}
The second condition comes from Eq.~(\ref{lag}):
\begin{align}
\begin{split}
\label{split}
  &{\rm Tr}_{AC}\left(\rho_{\mathrm{out},AC} \Pi^{\mathrm{fid}}\right)\\
\geq& \mathbb{E}_{\rho_{\mathrm{out},AC}}[\Lambda_{m,r}(|\hat{\omega}-(-1)^{\hat{a}}\beta|^2)]  \hspace{10pt} (m : {\rm odd} ),
\end{split}
\end{align}
where
\begin{equation}
\label{pifid}
 \Pi^{\mathrm{fid}} \coloneqq \ket{0}\bra{0}_{A} \otimes \ket{\beta}\bra{\beta}_{C} + \ket{1}\bra{1}_{A} \otimes \ket{-\beta}\bra{-\beta}_{C}.
\end{equation}
We note that the right-hand side of Eq.~(\ref{split}) corresponds
to $\hat{F}/(Np_{\rm test})$ in the asymptotic limit.

The analysis in the asymptotic limit now reduces to finding a bound on 
${\rm Tr}(M_{\rm ph} \rho_{\mathrm{out},AC} )$ under the constraints Eq.~(\ref{Pi_sig}) and Eq.~(\ref{split}).
 In Methods, we derive a family of bounds  $B(\kappa, \gamma)$ with nonnegative parameters  $\kappa$ and $\gamma$
satisfying 
\begin{align}
 {\rm Tr}\left(     \rho_{AC}  (M_{\rm ph} + \kappa \Pi^{\rm fid} - \gamma \Pi_{-}^{\rm\textcolor{black}{sig}}) \right) \le B(\kappa, \gamma)  \label{ast}
\end{align}
for any density operator $\rho_{AC}$.
Using  $B(\kappa, \gamma)$, an upper bound on the phase error rate in the asymptotic limit is given by
\begin{align}
\frac{\hat{N}_{\rm ph}}{N p_{\rm sig}}  \le u(\hat{F}/N p_{\rm test})\coloneqq B(\kappa, \gamma) - \kappa \hat{F}/N p_{\rm test} + \gamma q_- .
\end{align}

Next, we consider the finite-key analysis. 
From the asymptotic analysis, we expect that the bound $U(\hat{F})$ in (\ref{phase_ineq}) will be written in the form $U({\hat{F}})= N p_{\rm sig}  u(\hat{F}/N p_{\rm test}) + \Delta$, in which the finite-size correction $\Delta$ is to be determined. 
In order to find $\Delta$ that bounds $\hat{N}_{\rm ph}$ from above under general attacks, we use Azuma's inequality~\cite{CIT064}.
It is applicable to a series of events with general correlations,
and can be used to analyze a large deviation of the total sum $\sum_{i=1}^N \hat{T}^{(i)}$
of a series of random variables $\{{T}^{(i)}\}$ from its expectation if there is a known constraint on the 
conditional expectation of ${T}^{(i)}$, where conditioning is made on the events prior to the $i$-th.
In order to apply this inequality to our case,
we write $\hat{N}_{\rm ph}= \sum_{i=1}^N \hat{N}_{\rm ph}^{(i)}$ and $\hat{F}= \sum_{i=1}^N \hat{F}^{(i)}$ and 
seek a constraint on the conditional statistics of random variables $(\hat{N}_{\rm ph}^{(i)}, \hat{F}^{(i)})$.
A problem here is that the conditional state $\rho^{(i)}_{AC}$ on systems $A$ and $C$ has no guarantee on its reduced 
state on $A$, and hence does not satisfy ${\rm Tr}(\rho^{(i)}_{AC} \Pi_{-}^{\rm\textcolor{black}{sig}})=q_-$. This prevents us from using Eq.~(\ref{defq-}) to 
find a tight constraint on the statistics of $(\hat{N}_{\rm ph}^{(i)}, \hat{F}^{(i)})$.

One way to connect the property of Eq.~(\ref{defq-}) to Azuma's inequality is to add a measurement 
corresponding to the operator $\Pi_{-}^{\rm\textcolor{black}{sig}}$ to the protocol. 
In Ref.~\cite{m}, the trash round, in which Alice and Bob discards every data, is randomly chosen with a probability $p_{\rm trash}$
in Actual protocol.
This modification allows us to assume that, after Entanglement-sharing protocol, 
Alice makes measurement
 $\{ \Pi_{-}^{\rm\textcolor{black}{sig}}, \bm{1}-\Pi_{-}^{\rm\textcolor{black}{sig}}\}$ in the trash rounds to determine the total number $\hat{Q}_-$ of the events corresponding to $\Pi_{-}^{\rm\textcolor{black}{sig}}$.
Its purpose is to treat the property of Eq.~(\ref{defq-}) as that of a measurement outcome. 
In fact, the expectation of $\hat{Q}_-$ is $N p_{\rm trash} q_-$, and its deviation can be easily analyzed. Azuma's inequality is then used to analyze the large deviation of the three variables, $(\hat{N}_{\rm ph}, \hat{F}, \hat{Q}_-)$, for which the conditional statistics can be directly bounded using Eq.~(\ref{ast}). An obvious drawback in this approach is that it wastes $N p_{\rm trash}$ rounds in Actual protocol and lowers the finite-size key rate. 
%

Here we improve the above approach in such a way that one does not need to 
waste rounds. It comes from the observation 
that the operator $\Pi_{-}^{\rm\textcolor{black}{sig}}$ commutes with $M_{\rm ph}$.
It means that
we can perform the measurement $\{ \Pi_{-}^{\rm\textcolor{black}{sig}}, \bm{1}-\Pi_{-}^{\rm\textcolor{black}{sig}}\}$ at the
same time as the measurement of the phase error at the signal round.
Thus, we do not
have to add the trash rounds to 
obtain $\hat{Q}_-$. Since this structure is
in common with Ref.~\cite{m}, the same trick
is also applicable to the protocol of
Ref.~\cite{m}.
An explicit form of $U(\hat{F})$ and the proof of Eq.~(\ref{phase_ineq}) is given in Methods.


\section{Numerical simulation}
\label{section-simulation}

We simulated the key rate $G$ for the Gaussian channel
specified by a transmissivity $\eta$ and an excess noise $\xi$. 
The transmissivity $\eta$ represents the amplitude damping of a coherent state.
The excess noise $\xi$ represents an environmental noise generated on Bob's side, which increases the variance by a factor of $(1 + \xi)$ \cite{namiki2004practical,hirano2017implementation}
 (see Methods for the explicit definition).
We set $s=104, s'=51$, and $\epsilon=2^{-s}$.
It makes the protocol $2^{-50}$-secure.

For the predetermined parameters $(m,r)$ of the bounded function $\Lambda_{m,r}$, we adopt $(m,r) = (1,0.4120)$, which leads to $(\max\Lambda_{m,r},\min\Lambda_{m,r}) = (2.824,-0.9932)$ as shown in Ref.~\cite{m}.
As for $\beta$ and $f_{\rm suc,0}(x)$, we adopt $\beta = \sqrt{\eta\mu}$ and $f_{\rm suc,0}(x)=\Theta(x-x_{\rm th})$,
where $\Theta(x)=1$ for $x\geq 0$ and $\Theta(x)=0$ for $x<0$.
For each transmissivity $\eta$, we determined two coefficients $(\kappa , \gamma)$ via a convex optimization using the CVXPY 1.0.25 \cite{diamond2016cvxpy,agrawal2018rewriting} and  four parameters $(\mu , p_{\rm sig} , p_{\rm test} , x_{\rm th} )$ via the Nelder-Mead in the
scipy.minimize library in Python, in order to maximize the key rate.

\begin{figure}[b]
\centering
\includegraphics[width=9cm,angle=0,clip]{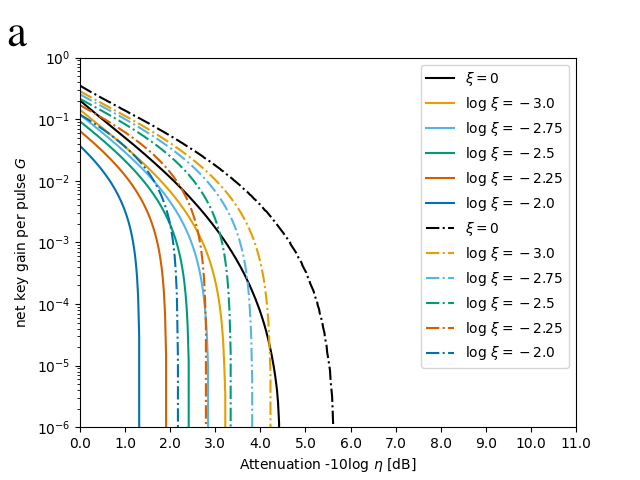}
\includegraphics[width=9cm,angle=0,clip]{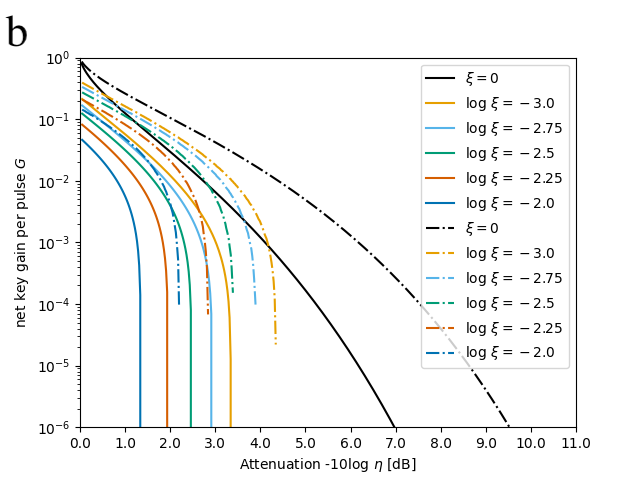}
\caption{Comparison of the secure key rates of our protocol and the homodyne protocol \cite{m}. 
The abscissa represents attenuation in decibel, i.e., $10 \log_{10}\eta$, where $\eta$ is the channel transmission.
Solid lines are the key rates of our protocol. Dash-dotted lines are those of the homodyne protocol.
{\bf a}: The finite-size key rates for various values of excess noise $\xi$ for $N = 10^{11}$ rounds.
{\bf b}: The asymptotic key rates for various values of $\xi$.
}
\label{fig1}
\end{figure}

We compare the key rate of our protocol with that in the previous protocol \cite{m} (we call it ``homodyne protocol'' henceforth)
in Fig.~\ref{fig1} at $\xi = 0 , 10^{-3.0} , 10^{-2.75} , 10^{-2.5} , 10^{-2.25} , 10^{-2.0}$
for $N =10^{11}$ and in the asymptotic limit. 
As expected, the key rate of our protocol using only the heterodyne measurement is lower than that of the homodyne protocol.
We see that if we shift each curve for the homodyne protocol by $\SI{-1}{\decibel}$, it will be close to the corresponding curve for our protocol, \textcolor{black}{except in the case of ($\xi = 0$ and $N\rightarrow\infty$)}.



\section{Discussion}

\begin{figure}[b]
\centering
\includegraphics[width=9cm,angle=0,clip]{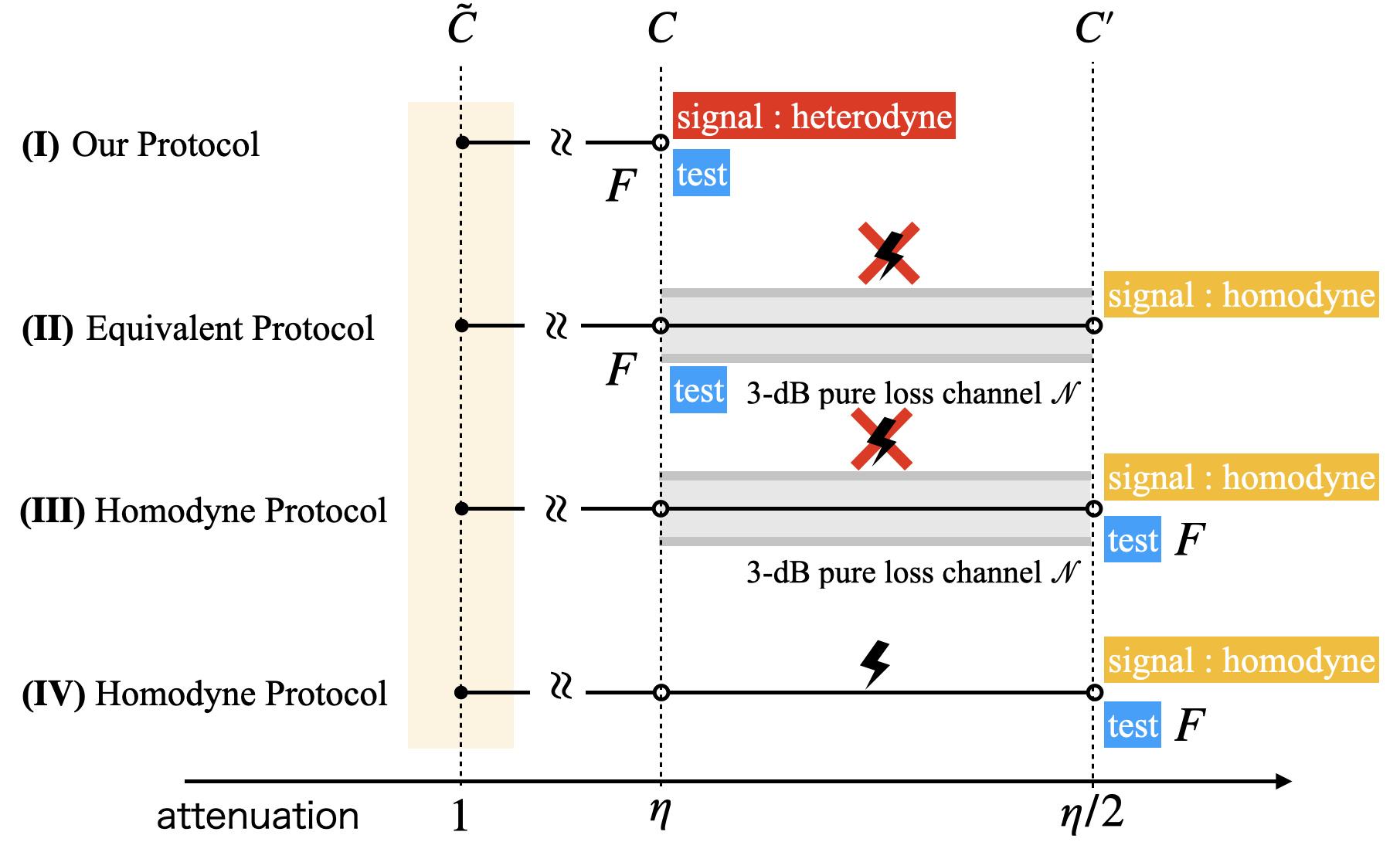}
\caption{Four protocols for understanding the difference between the proposed all-heterodyne protocol and the homodyne protocol: ($\1$) Our protocol, ($\2$) Equivalent protocol, ($\3$) Homodyne protocol with trusted loss channel, and ($\4$) Homodyne protocol with untrusted loss channel.}
\label{fig3}
\end{figure}


Since our protocol uses a binary phase modulation, the heterodyne measurement in the signal rounds
is used only for measuring the amplitude in one quadrature. This is also seen in Eqs. (\ref{pr1})--(\ref{pr3}),
where only $\omega_R$ is used and there is no reference to $\omega_I$. This observation may lead to 
an alternative way of proving its security as follows.

Although the main motivation in using the heterodyne measurement is to
dispense with the phase locking of the two LOs, for the sake of proving the security, 
one can assume that Bob uses an LO phase-locked to Alice's and  
configure his apparatus in Fig. \ref{fig0} such that
outcome $\omega_R$ is obtained from the upper-right pair of detectors, while outcome $\omega_I$ is
from the lower-left pair. Then, in a signal round, the lower part are redundant and the measurement
of $\omega_R$ is equivalent to a homodyne measurement placed behind a half beam splitter.
This suggests that one may just repurpose the security proof 
and the key length formula of the homodyne protocol \cite{m}, as it is, to our protocol.
In what follows, we argue that it is indeed true in the asymptotic limit but the 
achievable key rate is much worse than that of the security proof presented in the previous sections.

Let us introduce four protocols, summarized in Fig.~\ref{fig3}, by modifying our protocol in stages toward the homodyne protocol of Ref.~\cite{m}.

\begin{itemize}
  \item[($\1$)] (Our protocol) Bob performs the heterodyne measurement on the received pulse $C$ for the  signal and test.
  												Based on the test results, Alice and Bob confirm that the fidelity of the pulse $C$ to the coherent state $\ket{\beta}$ is no smaller than $F$.
  \item[($\2$)] (Equivalent protocol) 
							In the test round, Bob performs the heterodyne measurement on the received pulse $C$.
							Based on the test results, Alice and Bob confirm that the fidelity of the pulse $C$ is no smaller than $F$.
							In the signal round, Bob sends the received pulse $C$ to a $3$-dB-pure-loss channel which is out of Eve's reach. Then he performs the homodyne measurement on the output $C'$ of the $3$-dB-pure-loss channel. As described below, this protocol is equivalent  to ($\1$).
							
 \item[($\3$)] (Homodyne protocol with trusted loss channel)
 							The signal round is the same with the protocol ($\2$).
							In the test round, Bob performs the heterodyne measurement on $C'$.
							Based on the results, Alice and Bob confirm that
							the fidelity of the pulse $C'$ is no smaller than $F$.
	
	\item[($\4$)] (Homodyne protocol with untrusted loss channel)  This protocol itself is the same with the protocol ($\3$). In this case, unlike ($\2$) and ($\3$), we assume that Eve can attack the channel from $C$ to $C'$.
 \end{itemize}

Protocol ($\1$) is the all-heterodyne protocol proposed in this paper. 
Protocol ($\4$) is the case where the homodyne protocol of Ref.~\cite{m} is carried out 
on a channel whose transmission is lower by $3$ dB.
In the following, we compare the above four protocols with the same value of the fidelity bound $F$.

First, we compare ($\1$) and ($\2$).
As explained above, the heterodyne protocol ($\1$) is equivalent to a half beam splitter 
followed by a homodyne measurement. Since a half beam splitter is equivalent to 
a 3-dB-pure-loss channel, protocol ($\2$) is equivalent to ($\1$).

To compare other protocols, let us denote the $3$-dB-pure-loss channel appearing in protocols ($\2$) and ($\3$) by a CPTP (completely positive and trace preserving) map  $\mathcal{N}^{C\rightarrow C'}$,
 which satisfies
\begin{align}
\label{coh}
\mathcal{N}^{C\rightarrow C'}(\ket{\omega}\bra{\omega}) = \ket{\omega/\sqrt{2}} \bra{\omega/\sqrt{2}}
\end{align}
for any coherent state $\ket{\omega}$.
We compare the protocols ($\2$), ($\3$), and ($\4$) by considering Eve's possible attacks 
in each case, assuming that the same value of the fidelity bound $F$ was observed in the test. 
An attack by Eve can be characterized by a CPTP map
from $\tilde{C}$ to $C'E$, where $E$ means Eve's system. We denote the allowed sets of maps for the three protocols by $\mathcal{L}_{\2},\mathcal{L}_{\3}$, and $\mathcal{L}_{\4} $. 
To simplify the notation, we introduce an abbreviation
\begin{align}
 \overline{\mathcal{E}} \coloneqq \left( \mathcal{N}^{C\rightarrow C'}\otimes {\rm id}^{E} \right) \circ \mathcal{E}^{\tilde{C}\rightarrow CE},
\end{align}
and the density operator for the state of $\tilde{C}$ prepared by Alice as
\begin{align}
\rho_{\rm in}^a\coloneqq \ket{(-1)^a\sqrt{\mu}}\bra{(-1)^a\sqrt{\mu}}
\end{align}
Then, the sets $\mathcal{L}_{\2}$, $\mathcal{L}_{\3}$, and $\mathcal{L}_{\4}$ can be written as
\begin{align}
  \mathcal{L}_{\2}&= \Bigl\{ \: \overline{\mathcal{E}} \;\Big| \frac{1}{2}\sum_{a=0,1} \nonumber \\
 &\Braket{(-1)^a\sqrt{\eta\mu}| {\rm Tr}_E \mathcal{E}^{\tilde{C}\rightarrow CE}\left( \rho^a_{\rm in}   \right)|(-1)^a\sqrt{\eta\mu}} \ge 
 F  \Bigr\}, \\
 \mathcal{L}_{\3}&= \Bigl\{ \: \overline{\mathcal{E}} \;\Big|\frac{1}{2}\sum_{a=0,1} \\
 &\Braket{(-1)^a\sqrt{  \frac{\eta\mu}{2}   }| {\rm Tr}_E \overline{\mathcal{E}} 
  \left( \rho^a_{\rm in}   \right)|(-1)^a\sqrt{\frac{\eta\mu}{2}  }} \ge 
 F  \Bigr\}, \\
  \mathcal{L}_{\4}&= \Bigl\{ \: \mathcal{E}^{\tilde{C}\rightarrow C'E}  \;\Big| \frac{1}{2}\sum_{a=0,1} \nonumber \\
 & \Braket{(-1)^a\sqrt{ \frac{\eta\mu}{2} }| {{\rm Tr}_E} \mathcal{E}^{\tilde{C}\rightarrow C'E} \left( \rho^a_{\rm in}   \right)|(-1)^a\sqrt{\frac{\eta\mu}{2}}} \ge 
 F  \Bigr\}.
\end{align}

From the monotonicity of the fidelity and Eq.~(\ref{coh}), we find
\begin{align}
 \begin{split}
  \braket{(-1)^a\sqrt{\eta\mu}| {{\rm Tr}_E} \mathcal{E}^{\tilde{C}\rightarrow CE}\left( \rho^a_{\rm in}   \right)|(-1)^a\sqrt{\eta\mu}} \\
 \le  \braket{(-1)^a\sqrt{\eta\mu/2}| {{\rm Tr}_E} \overline{\mathcal{E}}\left( \rho^a_{\rm in}   \right)|(-1)^a\sqrt{\eta\mu/2}}.
  \end{split}
\end{align}
It means $\mathcal{L}_{\2} \subset\mathcal{L}_{\3}$.
Since the map $\bar{\mathcal{E}}$ is a special case of the general CPTP map $\mathcal{E}^{\tilde{C}\rightarrow C'E}$, $\mathcal{L}_{\3} \subset\mathcal{L}_{\4}$ holds.
We thus obtain 
\begin{align}
\label{subset}
 \mathcal{L}_{\2} \underset{ \rm(\hspace{.10em}i\hspace{.10em})  }{\subset} \mathcal{L}_{\3} \underset{ \rm(\hspace{.08em}ii\hspace{.08em}) }{\subset} \mathcal{L}_{\4} .
\end{align}

\begin{figure}[H]
\centering
\includegraphics[width=9cm,angle=0,clip]{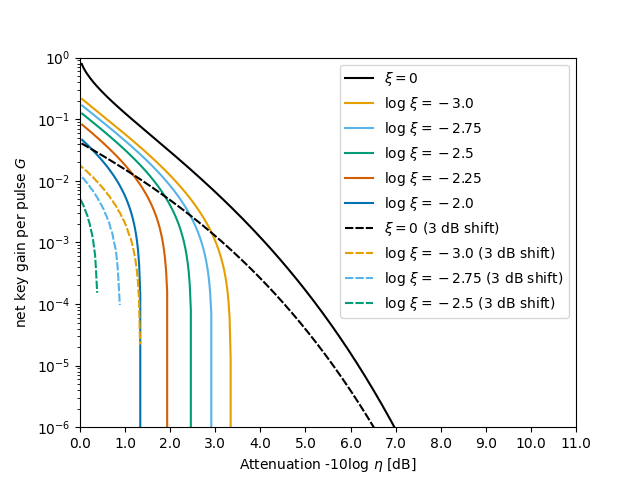}
\caption{Comparison of the secure key rates of the heterodyne protocol obtained from the two security arguments. 
Solid curves are the rates derived from our security proof. Broken curves are the rates obtained by 
repurposing the key-rate formula \cite{m} for the homodyne protocol. The latter curves are 
exactly the one shifted by $3$ dB from the rate curves of the homodyne protocol.}
\label{fig2}
\end{figure}

The above inclusive relation with the equivalence between Protocols ($\1$) and ($\2$)
justifies that we are able to repurpose the key rate formula for the homodyne protocol of 
Ref.~\cite{m} to achieve a secure key rate for the heterodyne protocol. In Fig.~\ref{fig2}, 
the key rates from such a repurposed formula are plotted as broken curves.
Because of the property of our channel model that the noise characteristics 
(see Eq.~(\ref{pxi})) are independent of the channel transmission, 
those curves are exactly the one shifted by $3$ dB from the rate curves of the 
homodyne protocol. On the other hand, as seen in Sec. \ref{section-simulation}, the key rate obtained from 
the present security proof (also plotted in Fig.~\ref{fig2}) has only about $1$-dB degradation. 

Since the removal of trash rounds does not affect the asymptotic key rate, we can 
ascribe the origin of the key rate difference between the two security proofs shown in 
Fig.~\ref{fig2} to two factors deduced from the inclusive relation of Eq.~(\ref{subset}): $\rm(\hspace{.18em}i\hspace{.18em})$ the repurposed formula 
assumes the fidelity test at a different point and may fail to fully utilize the observed fidelity 
bound $F$ and $\rm(\hspace{.08em}ii\hspace{.08em})$ the repurposed formula overestimates Eve's ability as if she could eavesdrop on the 
fictitious $3$-dB loss channel.

In conclusion, by 
using only heterodyne measurement, our protocol eliminates the need for the phase locking of the local oscillators used by the sender and the receiver.
It thus makes its implementation easier and enhances the practical advantages of the CV-QKD protocol. 
Comparison with a similar protocol with a homodyne measurement and with the same level of the security, 
our all-heterodyne protocol suffers only $1$-dB penalty in the rate-distance curve. This is much better than 
the naive expectation based on the fact that a heterodyne detection in one quadrature is equivalent to 
a homodyne measurement with a $3$-dB loss.
In addition, we improved the  proof technique to remove the ``trash rounds'' required in the homodyne protocol \cite{m}, which simplifies the structure of the protocol.

\section*{Methods}
\subsection{Derivation of operator bound  $B(\kappa,\gamma)$}
\label{Phase error bound}
Here we construct a computable bound $B(\kappa,\gamma)$, which 
satisfies an operator inequality
\begin{align}
M[\kappa, \gamma]
&\coloneqq M_{\rm ph}  + \kappa \Pi^{\rm fid}  -\gamma \Pi_{-}^{\rm\textcolor{black}{sig}} \label{Mevalue} \\
&\le B(\kappa, \gamma) \bm{1} 
\label{astarisk}
\end{align}
for $\kappa \ge 0$ and $\gamma \ge 0$.
Note that Eq.~(\ref{astarisk}) implies that
\begin{align}
 {\rm Tr}\left(M[\kappa, \gamma] \rho_{AC}\right)
\le B(\kappa, \gamma)
\end{align}
for any density operator $\rho_{AC}$.

We denote by $\sigma_{\sup}(O)$ the supremum of the spectrum of a bounded self-adjoint operator $O$.
The following lemma is shown in Ref.~\cite{m}.
\begin{lem}
\label{lemma1}
Let $\Pi_{\pm}$ be orthogonal projections satisfying $\Pi_{+}\Pi_{-} = 0$. Suppose that the rank of $\Pi_{\pm}$ is no smaller than 2 or infinite. Let $M_{\pm}$ be self-adjoint operators satisfying $\Pi_{\pm} M_{\pm} \Pi_{\pm} = M_{\pm} \leq \alpha_{\pm}\Pi_{\pm}$ ,where $\alpha_{\pm}$ is real constants. Let $\ket{\psi}$  be an unnormalized vector satisfying $(\Pi_{+}+\Pi_{-})\ket{\psi} = \ket{\psi}$ and $\Pi_{\pm}\ket{\psi} \neq 0$.
Define the following quantities with respect to $\ket{\psi}:$
\begin{align}
C_{\pm} & \coloneqq \bra{\psi} \Pi_{\pm} \ket{\psi} (>0)\\
D_{\pm} & \coloneqq C^{-1}_{\pm}  \bra{\psi} M_{\pm} \ket{\psi} \\
V_{\pm} & \coloneqq C^{-1}_{\pm}  \bra{\psi} M^2_{\pm} \ket{\psi} - D^2_{\pm} .
\end{align}
Then, for any real numbers $\gamma_{+},\gamma_{-}$, we have
\begin{align}
\sigma_{\rm sup}\left(  M_{+} + M_{-} + \ket{\psi}\bra{\psi} -\gamma_{+}\Pi_{+} -\gamma_{-} \Pi_{-}   \right) \\
\leq\sigma_{\rm sup}\left(  M_{4d}   \right),
\end{align}
where $M_{4d}$ is defined as
\begin{align}
&M_{\rm 4d}\coloneqq \nonumber \\
&\left[
    \begin{array}{cccc}
    \alpha_{+}-\gamma_{+} & \sqrt{V_{+}} & 0 & 0 \\
    \sqrt{V_{+}} & C_{+} + D_{+} - \gamma_{+} & \sqrt{C_{+}C_{-}} & 0 \\
    0 & \sqrt{C_{+}C_{-}} & C_{-} + D_{-} - \gamma_{-} & \sqrt{V_{-}} \\
    0 & 0 & \sqrt{V_{-}} & \alpha_{-} - \gamma_{-} \\
    \end{array}
\right].
\end{align}
\end{lem}
In order to apply this lemma to our case, 
we first derive an explicit form of $M_{\rm ev(od)}$. Let 
$\Pi_{\rm ev}$ (resp. $\Pi_{\rm od}$ ) be the projection to the subspace with even (resp. odd) photon numbers, i.e., $\Pi_{\rm ev} + \Pi_{\rm od} = {\bm 1}$ and 
$(\Pi_{\rm ev} - \Pi_{\rm od})\ket{\omega} = \ket{-\omega} $.
Rewriting Eq.~(\ref{actCP}) in the $X$ basis leads to
\begin{align}
\label{Eq:komega1}
 K(\omega) = \sqrt{\frac{2f_{{\rm suc},0}(\omega_R)}{\pi}}\left(  \ket{+}_B\bra{\omega}_C\Pi_{\rm ev } +\ket{-}_B\bra{\omega}_C\Pi_{\rm od}     \right).
\end{align}
Then we have
 \begin{align}
M_{\rm ev(od)} &=\int_{\mathbb{C}} d^2 \omega K^\dagger(\omega)\ket{+(-)}\bra{+(-)}_BK(\omega)\label{Eq:komega2}\\
&= \frac{2}{\pi} \int_{\mathbb{C}} d^2 \omega f_{\rm suc,0}(\omega_R) \Pi_{\rm ev(od)} \ket{\omega}\bra{\omega} \Pi_{\rm ev(od)}\label{Eq:komega3}.
\end{align}
We introduce
\begin{equation}
\begin{split}
\label{aaa3}
\ket{\phi_{\rm err}} \coloneqq &\ket{+}\otimes \Pi_{\rm od}\ket{\beta} + \ket{-}\otimes \Pi_{\rm ev}\ket{\beta},\\
\ket{\phi_{\rm cor}} \coloneqq &\ket{+}\otimes \Pi_{\rm ev}\ket{\beta} + \ket{-}\otimes \Pi_{\rm od}\ket{\beta},\\
M^{\rm err}[\kappa , \gamma] \coloneqq &\ket{+}\bra{+} \otimes M_{\rm od}  + \ket{-}\bra{-}\otimes M_{\rm ev}, \\
 &+ \kappa \ket{\phi_{\rm err}}\bra{\phi_{\rm err}} - \gamma \ket{-}\bra{-} \otimes \Pi_{\rm ev}, \\
  M^{\rm cor}[\kappa , \gamma] \coloneqq & \kappa \ket{\phi_{\rm cor}}\bra{\phi_{\rm cor}} -  \gamma \ket{-}\bra{-} \otimes \Pi_{\rm od}.
\end{split}
\end{equation}
Compared with Eq.~(\ref{pifid}), we see that the following relation holds:
\begin{align}
\label{aaa4}
 \Pi^{\rm fid} = \ket{\phi_{\rm err}}\bra{\phi_{\rm err}} + \ket{\phi_{\rm cor}} \bra{\phi_{\rm cor}}.
\end{align}
From Eqs.~(\ref{aaa1}), (\ref{aaa2}), (\ref{Pi_sig}), (\ref{Mevalue}), (\ref{aaa3}), (\ref{aaa4}),
we can then decompose $M[\kappa , \gamma] = M_{\rm ph}  + \kappa \Pi^{\rm fid}  -\gamma \Pi_{-}^{\rm\textcolor{black}{sig}} $ into a direct sum as
\begin{equation}
 M[\kappa , \gamma]  =  M^{\rm err}[\kappa , \gamma] \oplus M^{\rm cor}[\kappa , \gamma] \label{directsum}.
\end{equation}
We apply Lemma \ref{lemma1} to $M^{\rm err}[\kappa , \gamma]$ by choosing 
\begin{align}
M_{\pm} &= \ket{\pm}\bra{\pm}\otimes M_{\rm od(ev)} \\
\ket{\psi} &= \sqrt{\kappa}\ket{\phi_{\rm err}} \\
\Pi_{\pm} &=\ket{\pm}\bra{\pm}\otimes\Pi_{\rm od(ev)}\\
\alpha_{\pm} &= 1\\
\gamma_{+} &= 0 , \gamma_{-} = \gamma.
\end{align}
This leads to
\begin{align}
\sigma_{\rm sup}( M^{\rm err}[ \kappa,\gamma  ]) \le 
\sigma_{\rm sup} (M_{4d}^{\rm err}[ \kappa ,\gamma ]) \label{suprelation} ,  
\end{align} 
where
\begin{align}
\label{Eq:4derr}
&M_{\rm 4d}^{\rm err}[\kappa,\gamma]\coloneqq \nonumber \\
&\left[
    \begin{array}{cccc}
    1                          & \sqrt{V_{\rm od}}                & 0                                                    & 0 \\
    \sqrt{V_{\rm od}}  & \kappa C_{\rm od} + D_{\rm od}     &\kappa \sqrt{C_{\rm od}C_{ \rm ev }}         & 0 \\
    0                          & \kappa\sqrt{C_{\rm od}C_{\rm ev}} & \kappa C_{\rm ev} + D_{\rm ev} - \gamma & \sqrt{V_{\rm ev}} \\
    0                          & 0                                         & \sqrt{V_{\rm ev}}                            & 1 - \gamma \\
    \end{array}
\right] 
\end{align}
with
\begin{align}
C_{\rm ev} &\coloneqq  \bra{\beta} \Pi_{\rm ev} \ket{\beta} = e^{-|\beta|^2}\cosh |\beta|^2,   \label{Eq:Cev}  \\
C_{\rm od} &\coloneqq  \bra{\beta} \Pi_{\rm od} \ket{\beta} = e^{-|\beta|^2}\sinh |\beta|^2,  \label{Eq:Cod}  \\
D_{\rm ev(od)}  &\coloneqq   C^{-1}_{\rm ev(od)}  \bra{\beta} M_{\rm ev(od)} \ket{\beta},   \label{Eq:Devod} \\
V_{\rm ev(od)}  &\coloneqq   C^{-1}_{\rm ev(od)}  \bra{\beta}  \left( M_{\rm ev(od)} \right)^2 \ket{\beta} - D^2_{\rm ev(od)}.   \label{Eq:Vevod} 
\end{align}
Similarly, we apply Lemma \ref{lemma1} to $M^{\rm cor}[\kappa , \gamma]$ by choosing 
\begin{align}
M_{\pm} &= 0 \\
\ket{\psi} &= \sqrt{\kappa}\ket{\phi_{\rm cor}} \\
\Pi_{\pm} &=\ket{\pm}\bra{\pm}\otimes\Pi_{\rm ev(od)}\\
\alpha_{\pm} &= 0\\
\gamma_{+} &= 0 , \gamma_{-} = \gamma.
\end{align}
This leads to
\begin{align}
\sigma_{\rm sup}( M^{\rm cor}[ \kappa,\gamma  ]) \le 
\sigma_{\rm sup} (M_{4d}^{\rm cor}[ \kappa ,\gamma ]) \label{suprelation2} ,  
\end{align} 
where
\begin{align}
M_{\rm 4d}^{\rm cor}[\kappa,\gamma]\coloneqq 
\left[
    \begin{array}{cccc}
    0                          & 0           & 0                                                    & 0 \\
   0                           & \kappa C_{\rm ev}            &\kappa \sqrt{C_{\rm od}C_{ \rm ev }}         & 0 \\
    0                          & \kappa\sqrt{C_{\rm od}C_{\rm ev}} & \kappa C_{\rm od} - \gamma & 0 \\
    0                          & 0                                         &0                           &  - \gamma \\
    \end{array}
\right]. 
\end{align}
Next we define one of its block matrices by 
\begin{align}
\label{Eq:cor2d}
M_{\rm 2d}^{\rm cor}[\kappa,\gamma]\coloneqq 
\left[
    \begin{array}{cc}
    \kappa C_{\rm ev}            &\kappa \sqrt{C_{\rm od}C_{ \rm ev }}       \\
     \kappa\sqrt{C_{\rm od}C_{\rm ev}} & \kappa C_{\rm od} - \gamma \\
    \end{array}
\right],
\end{align}
which then satisfies
\begin{align}
\label{sigma4d}
\sigma_{\rm sup} (M_{4d}^{\rm cor}[ \kappa ,\gamma ]) = \max\{ \sigma_{\rm sup} (M_{2d}^{\rm cor}[ \kappa ,\gamma ]) , 0, -\gamma \}. 
\end{align} 
Since $\gamma, \kappa>0$, we have $\det \left( M_{2d}^{\rm cor}[ \kappa ,\gamma ] \right)= -\gamma \kappa C_{\rm ev} < 0 $ and hence $\sigma_{\rm sup} (M_{2d}^{\rm cor}[ \kappa ,\gamma ])>0$. We can thus simplify Eq. (\ref{sigma4d}) as 
\begin{align}
\sigma_{\rm sup} (M_{4d}^{\rm cor}[ \kappa ,\gamma ]) = \sigma_{\rm sup} (M_{2d}^{\rm cor}[ \kappa ,\gamma ]) . 
\label{suprelation3}
\end{align}
Then, from Eqs.~(\ref{directsum}), (\ref{suprelation}), (\ref{suprelation2}), and (\ref{suprelation3}), we finally obtain an upper bound $B(\kappa,\gamma)$ as
\begin{align}
\label{form-B}
 B(\kappa,\gamma) = \max\left(  \sigma_{\rm sup}\left(M^{\rm err}_{4d}[\kappa , \gamma]     \right)   ,\sigma_{\rm sup}\left(  M^{\rm cor}_{2d}[\kappa , \gamma]   \right)  \right),
\end{align}
which satisfies Eq.~(\ref{astarisk}).

\subsection{Detailed security proof}

Here, we
construct a function $U$ satisfying Eq.~(\ref{phase_ineq}),
which determines the final key length through Eq.~(\ref{final-key-length}) and guarantees the security. 

For that purpose, 
we will define a protocol which we call the estimation protocol.  It reproduces the statistics of 
$(\hat{N}_{\rm ph}, \hat{F})$ and is suited to the use of Azuma's inequality. The main difference from Entanglement-sharing protocol followed by the $X$-basis measurements is that Alice conducts the $X$-basis measurement on her qubit $A$
even when Bob's measurement outcome is a failure. We thus define the operators for such measurements as 
\begin{align}
M_{x_A, {\rm fail}}\coloneqq \ket{x_A}\bra{x_A} \otimes M_{\rm fail}
\end{align}
for $x_A=+,-$. 
The protcol is then formally defined as follows.
\\
\\
{\bf Estimation protocol}
\\
1.\ Alice prepares a qubit $A$ and an optical pulse $\tilde{C}$  in the state $\ket{\Psi}_{A\tilde C} $ defined in Eq.~(\ref{psiA}) and sends $\tilde C$ to Bob. She repeats it $N$ times.
\\
\\
2.\ For each of the $N$ rounds, with the probabilities $p_{\rm sig}$ and $p_{\rm test}$, Bob determines whether each round is ``signal'' or ``test'' and announces it.
Based on this label, Alice and Bob proceed as follows.
\\
\\
\lbrack\,signal\,\rbrack\,Alice and Bob measure their systems and obtain $(\hat{x}_A,\hat{x}_B)$, where the POVM elements are given  by $\{M_{x_A,x_B} \}_{x_A\in\{+,-\},x_B\in\{\mathrm{+,-,fail}\}}$ defined in Eq.~(\ref{aaa1}).
\\
\\
\lbrack\,test\,\rbrack\,Alice measures her qubit $A$ on the $Z$ basis (\{$\ket{0}, \ket{1}$\}) to obtain a bit $\hat{a}$. Bob performs a heterodyne measurement to obtain a complex number $\hat{\omega}$. 
 \\
 \\
 3.\ For $i=1, \dots , N$, variables $\hat{N}^{(i)}_{\rm ph}$,$\hat{F}^{(i)}$,$\hat{Q}^{(i)}_{-}$, and $\hat{T}^{(i)}$ are defined according to Table \ref{tableRD1} by using the outcomes in Step 2 for the $i$-th round.
Finally, the sum of these variables are defined as
 \begin{align}
  \hat{N}_{\rm ph}&= \sum_{i=1}^{N}\hat{N}^{(i)}_{\rm ph},\\
  \hat{F}&=\sum_{i=1}^{N}\hat{F}^{(i)},\\
  \hat{Q}_{-}&=\sum_{i=1}^{N}\hat{Q}^{(i)}_{-},\\
  \hat{T} &=\sum_{i=1}^N \hat{T}^{(i)}. 
 \end{align}
 
The way to determine $(\hat{N}_{\rm ph},\hat{F})$ is equivalent to Entanglement-sharing protocol followed by the $X$-basis measurement.
Therefore, if we can show that Eq.~(\ref{phase_ineq}) holds true in Estimation protocol, the security of Actual protocol immediately follows.


\begin{table*}
  \begin{tabular}{c|c|c|c|c|c} \hline
    round &outcome & $ \hat{N}^{(i)}_{\rm ph}$ & $\hat{F}^{(i)}$ & $\hat{Q}^{(i)}_{-}$ &$\hat{T}^{(i)}$\\
    \hline\hline
    \multirow{6}{*}{signal}      & $(\hat{x}_A,\hat{x}_B) = (+,+)$          & 0 &                                                  0&0&0                                                                       \\ \cline{2-6}
                                 & $(\hat{x}_A,\hat{x}_B)= (+,-)$          & 1 &                                                  0&0&$p_{\rm sig}^{-1}$                                                      \\ \cline{2-6}
                                 & $(\hat{x}_A,\hat{x}_B)= (-,+)$          & 1 &                                                  0&1&$(1-\gamma)/p_{\rm sig}$                                                \\ \cline{2-6}
                                 & $(\hat{x}_A,\hat{x}_B)= (-,-)$          & 0 &                                                  0&1&$-\gamma/p_{\rm sig}$                                                   \\ \cline{2-6}
                                 & $(\hat{x}_A,\hat{x}_B) = (+,{\rm fail})$ & 0 &                                                  0&0&0                                                                       \\ \cline{2-6}
                                 &$(\hat{x}_A,\hat{x}_B) = (-,{\rm fail})$  & 0 &                                                  0&1&$-\gamma/p_{\rm sig}$                                                   \\ \cline{1-6}
    test                         &$(\hat{a},\hat{\omega})$                  & 0 &$\Lambda_{m,r}(|\hat{\omega} - (-1)^{\hat{a}}\beta|^2)$&0&$\kappa \Lambda_{m,r}(|\hat{\omega} - (-1)^{\hat{a}}\beta|^2) /p_{\rm test}$\\ \cline{1-6}
  \end{tabular}
  \caption{\label{tableRD1} Measurement outcomes in Estimation protocol and definitions of the random variables.}
  \label{tb:mulrow}
\end{table*}

In contrast to Ref.~\cite{m}, here we defined $\hat{Q}_{-}$ without introducing the trash rounds.
It is achieved because $\hat{Q}^{(i)}_-$ can be simultaneously measured with $\hat{N}_{\rm ph}^{(i)}$ 
in Estimation protocol, which can be seen from the commutativity of the corresponding POVMs $\Pi_{-}^{\rm\textcolor{black}{sig}}$ and $M_{\rm ph}$. 
Thus, we can dispense with the trash rounds and improve the finite-key performance.

To find an upper bound $U(\hat{F})$ satisfying Eq.~(\ref{phase_ineq}),
we first find an upper bound on the expectation of $\hat{T}^{(i)}$ for arbitrary
state $\rho_{AC}$ on Alice's qubit $A$ and Bob's received pulse $C$.
From Table \ref{tableRD1}, we see that
$\hat{T}^{(i)}$ and $\hat{T}$ are related to other variables as
\begin{align}
  \hat{T}^{(i)} &= \frac{\hat{N}^{(i)}_{\rm ph}}{p_{\rm sig}} + \kappa \frac{\hat{F}^{(i)}}{p_{\rm test}} - \gamma\frac{\hat Q^{(i)}_-}{p_{\rm sig}},  \label{Tidef}\\
   \hat{T}&=  \frac{\hat{N}_{\rm ph}}{p_{\rm sig}} + \kappa \frac{\hat{F}}{p_{\rm test}} - \gamma\frac{\hat Q_-}{p_{\rm sig}}   \label{Tdef} .
\end{align}
For state $\rho_{AC}$, we have
%
%
%
\begin{align}
  \mathbb{E}_{\rho_{AC}}\left[ \hat{N}_{\rm ph}^{(i)} \right] = p_{\rm sig} {\rm Tr}\left(  M_{\rm ph}  \rho_{AC} \right)
  \label{N_ph_i}
\end{align}
and
\begin{align}
 \mathbb{E}_{\rho_{AC}}\left[ \hat{Q}_-^{(i)} \right] \label{Q_i}
=&p_{\rm sig} {\rm Tr}\left( \left( M_{-,+}+ M_{-,-}+ M_{-,\mathrm{fail}}\right)  \rho_{AC} \right) \nonumber\\
  =&p_{\rm sig} {\rm Tr}\left(  \Pi_{-}^{\rm\textcolor{black}{sig}}  \rho_{AC} \right).
\end{align}
According to Eq.~(\ref{lag}), the operator $\Pi^{\rm fid}$ satisfies
\begin{align}
\label{Fevalue}
\mathbb{E}_{\rho_{AC}} \left[ \hat{F}^{(i)} \right] \leq p_{\rm test} {\rm Tr}\left(  \Pi^{\rm fid} \rho_{AC} \right).
\end{align}
From the relations Eqs.~(\ref{ast}), (\ref{Tidef}), (\ref{N_ph_i}), (\ref{Q_i}) and (\ref{Fevalue}), we have 
\begin{align}
\label{bound-Ti}
 \mathbb{E}_{\rho_{AC}} \left[  \hat{T}^{(i)}   \right] &\leq  {\rm Tr}\left( \left( M_{\rm ph}  + \kappa \Pi^{\rm fid}  -\gamma \Pi_{-}^{\rm\textcolor{black}{sig}}  \right) \rho_{AC}  \right)\nonumber\\
 &\leq  B(\kappa,\gamma)
\end{align}
for any state $\rho_{AC}$.
Using this property, we can derive a bound on $\hat{T}$ in the form of
\begin{equation}
\label{bound-T-intuitive}
 \mathrm{Pr}\left[ \hat{T}-NB(\kappa,\gamma)\leq \delta_1(\epsilon/2) \right] \geq 1-\frac{\epsilon}{2}
\end{equation}
by using Azuma's inequality \cite{CIT064}. The detail is given in the next subsection and  $\delta_1(\epsilon)$ is defined in Eq.~(\ref{delta1}).

Since the variables $\{ \hat{Q}^{(i)}_- \}_i$ 
are outcomes on Alice's qubits, they are not affected by Eve's attack. 
From the initial state (\ref{psiA}), we see that they are $N$ independent 
Bernoulli trials. As a result, $\hat{Q}_-$ follows the binomial distribution 
with probability $p_{\mathrm{sig}}q_-$,
where $q_-$ is defined in Eq. (\ref{Eq_q}).
We may then derive a bound in the form of 
\begin{equation}
 \label{bound-Q-intuitive}
 \mathrm{Pr}\left[ \hat{Q}_- - Np_{\mathrm{sig}}q_- \leq \delta_2(\epsilon)\right] \geq 1-\frac{\epsilon}{2}
\end{equation}
by using the Chernoff-Hoeffding bound \cite{hoeffding1994probability}. The detail is given in the next subsection and  $\delta_2(\epsilon)$ is defined in Eq.~(\ref{delta2}).

Combining Eqs.~(\ref{Tdef}), (\ref{bound-T-intuitive}) and (\ref{bound-Q-intuitive}), 
we obtain an explicit form of $U(\hat{F})$ as
\begin{align}
\label{form-of-U}
U(\hat{F}) 
=& -\kappa \frac{p_{\rm sig}}{p_{\rm test}}\hat{F} + \gamma\left(  Np_{\rm sig}q_- + \delta_2\left(\frac{\varepsilon}{2}\right)   \right) \nonumber \\
	&+ p_{\rm sig}\left( NB(\kappa,\gamma) + \delta_1\left(\frac{\varepsilon}{2}\right) \right),
\end{align}
which satisfies
\begin{equation}
{\rm Pr}\left[ \hat{N}_{\rm ph} \leq U(\hat{F}) \right] \geq 1-\varepsilon.
\end{equation}
This formula refers to Eqs.~\eqref{Eq_q}, \eqref{Eq:komega1}--\eqref{Eq:komega3}, \eqref{Eq:4derr}--\eqref{Eq:Vevod}, \eqref{Eq:cor2d}, \eqref{form-B}, \eqref{Eq:cmin}--\eqref{delta1} and \eqref{delta2} for the definitions used in it.

\subsection{Derivation of finite-size corrections 
\\$\delta_1(\epsilon)$ and $\delta_2(\epsilon)$}
Here we derive explicit forms of $\delta_1(\epsilon)$ and $\delta_2(\epsilon)$ appearing in Eqs.~(\ref{bound-T-intuitive}) and (\ref{bound-Q-intuitive}). For $\delta_1 (\epsilon)$,  
we utilize Azuma's inequality \cite{CIT064} in the form of the following proposition:
\begin{prop}{\bf Azuma's inequality}
\label{Az}
Suppose that $( \hat{X}^{(k)} )_{k =0,1, \ldots}$ is a martingale  and  $( \hat{Y}^{(k)} )_{k =1, 2, \ldots}$ is a predictable process with regard to $( \hat{X}^{(k)} )_{k =0, 1, \ldots}$, which satisfies
\begin{align}
\label{cbound}
 -\hat{Y}^{(k)} + c_{\rm min} \leq   \hat{X}^{(k)} - \hat{X}^{(k-1)}\leq -\hat{Y}^{(k)} + c_{\rm max}
\end{align}
for constants $c_{\rm min} $ and $c_{\rm max}$.
Then for all positive integers N and all positive reals $\delta$, 
\begin{align}
{\rm Pr}[\hat{X}^{(N)}-\hat{X}^{(0)}\geq \delta] \leq \exp  \left(-\frac{ 2\delta^2 }
{ (c_{\rm max} - c_{\rm min})^2 N } \right).
\end{align}
\end{prop}
Here, we say a sequence $(\hat{Y}^{(k)})_{k =1, 2, \ldots}$ is a  predictable process with respect to a sequence $(\hat{X}^{(k)})_{k =0, 1, \ldots}$ when 
$\mathbb{E}[\hat{Y}^{(k)}|\hat{X}^{<k}] =\hat{Y}^{(k)} $ for all  $k\ge 1$, where $\hat{X}^{<k} \coloneqq ( \hat{X}^{(0)} ,\hat{X}^{(1)} ,\dots,\hat{X}^{(k-1)}  )$.
To apply Azuma's inequality for $\hat{T}$, we use Doob decomposition of $( T^{(k)} )_{k=1,\dots, N}$, given by
\begin{align}
\hat{X}^{(0)}  & =  0, \\
\hat{X}^{(k)} & = \sum_{i =1}^{k} (\hat{T}^{(i)} - \hat{Y}^{(i)} ), \hspace{1cm}  k\geq1, \\
\hat{Y}^{(i)} & =  \mathbb{E}[\hat{T}^{(i)}|\hat{X}^{< i}] \label{yi}.
\end{align}
This definition guarantees that $( \hat{X}^{(k)} )_{k =0, 1, \ldots, N}$ is a martingale, and $( \hat{Y}^{(k)})_{k =1, 2, \ldots, N}$ is a predictable process.
According to Table \ref{tableRD1}, 
$\hat{T}^{(i)}$ satisfies 
\begin{align}
 c_{\rm min} \le \hat{T}^{(i)} \le c_{\rm max}
\end{align}
with
\begin{align}
 c_{\rm min} &=  \min( p_{\rm test}^{-1}\kappa\min_{\nu\ge 0}\Lambda_{m,r}(\nu) , -p^{-1}_{\rm sig}\gamma ),\label{Eq:cmin}\\
 c_{\rm max} &= \max( p_{\rm test}^{-1}\kappa\max_{\nu\ge 0}\Lambda_{m,r}(\nu) , p^{-1}_{\rm sig}),\label{Eq:cmax}
\end{align}
and hence this choice fulfills Eq.~(\ref{cbound}).
We define
\begin{align}
\label{delta1} 
\delta_1(\varepsilon) \coloneqq \left( c_{\rm max}- c_{\rm min} \right) \sqrt{\frac{N}{2}\ln\left( \frac{1}{\varepsilon} \right)}.
\end{align}
By setting $\delta = \delta_1\left( \varepsilon /2 \right)$ in Proposition \ref{Az}, we obtain
\begin{align}
 {\rm Pr}\left[\hat{T}[\kappa,\gamma] \leq \sum_{i=1}^{N}\hat{Y}^{(i)} +  \delta_1\left( \varepsilon /2 \right) \right] \geq 1 - \frac{\varepsilon}{2}. 
\end{align}
Since Eqs.~(\ref{bound-Ti}) and (\ref{yi}) imply $\hat{Y}^{(i)} \leq B(\kappa, \gamma)$,
we obtain Eq.~(\ref{bound-T-intuitive}).

Next, we derive an explicit form of $\delta_2(\epsilon)$. 
Since the sequence of outcomes $\{\hat{Q}_-^{(i)}\}_i$ obeys the Bernoulli distribution with probability $p_{\mathrm{sig}}q_-$,
we can use the Chernoff-Hoeffding bound \cite{hoeffding1994probability} to obtain
\begin{align}
\begin{split}
\label{chernoffq-}
 &{\rm Pr }\left[ \hat{Q}_- \geq Np_{\rm sig}q_- +\delta  \right] \\
\leq &\exp\left[ -ND \left( p_{\rm sig}q_- + \frac{\delta}{N} \middle|
\middle| p_{\rm sig}q_ -\right)   \right] 
\end{split}
\end{align}
with $0<\delta<(1-p_{\rm sig}q_-)N$, where
\begin{align}
D\left( x||y    \right)  \coloneqq x\log\frac{x}{y} + \left(  1 - x   \right)\log\frac{1-x}{1-y}
\end{align} 
is the Kullback-Leibler divergence.
We define $\delta_2(\varepsilon) > 0$ as the unique solution for the following equations:
\begin{align}
\begin{cases}
\label{delta2}
D \left( p_{\rm sig}q_- + \frac{\delta_2(\varepsilon)}{N} \middle|\middle| p_{\rm sig}q_ -\right)  =\frac{-\log\varepsilon}{N} & \left( \varepsilon > (p_{\rm sig}q_-)^N \right)\\
\delta_2(\varepsilon) = (1-p_{\rm sig}q_-)N & \left( \varepsilon \le (p_{\rm sig}q_-)^N \right).
\end{cases} 
\end{align}
Then, combined with Eq.~(\ref{chernoffq-}),  we have 
\begin{align}
 {\rm Pr}\left[  \hat{Q}_- \leq Np_{\rm sig}q_-  + \delta_2(\varepsilon)  \right] \geq 1 - \varepsilon ,
 \label{bound2}
\end{align}
from which Eq.~(\ref{bound-Q-intuitive}) follows.

\subsection{Channel model and simulation}
Calculation of the final key length $\hat{N}^{\mathrm{fin}}$ 
in Sec.~\ref{section-simulation}
was done by assuming a channel model, from which we determined the values of 
parameters  $\hat{N}^{\mathrm{suc}},\hat{F}$, and $H_{\mathrm{EC}}$.
We adopted a Gaussian channel as a model, and here we describe its detail. 

Our model is characterized by transmissivity $\eta$ and excess noise $\xi$.
When Alice sends $\ket{(-1)^a\sqrt{\mu}}$ through this channel, the state $\rho^a_{\rm model}$ that Bob receives is written as
\begin{equation}
 \rho_{\rm model}^a  \coloneqq \int_{\mathbb{C}} d^2 \gamma p_{\xi}(\gamma)\ket{(-1)^a\sqrt{\eta\mu}+\gamma}\bra{  (-1)^a\sqrt{\eta\mu}+\gamma},
\end{equation}
with 
\begin{align}
\label{pxi}
 p_{\xi}(\gamma) \coloneqq \frac{ 2 }{ \pi \xi }e^{-2|\gamma|^2/\xi}.
\end{align}

Under this channel model, the expectation value of $\hat{F}$ is calculated as
\begin{equation}
\begin{split}
   \mathbb{E}\left[   \hat F  \right] &= Np_{\rm test}\mathbb{E}\left[   \Lambda_{m,r}\left(  |\hat\omega - (-1)^a\sqrt{\eta\mu}|^2   \right)  \right] \\
   &=\frac{p_{\rm test}N}{1+\xi/2}\left[ 1-(-1)^{m+1}\left(  \frac{\xi/2}{1+r(1+\xi/2)}   \right)^{m+1}    \right]
\end{split}
\end{equation}
We used this value as the observed value of $\hat{F}$ in Actual protocol.

For $\hat{N}^{\rm suc}$,
let us define the probability $P^{\pm}$ that Alice and Bob succeed in the detection and have the same/different bits in the signal round in Actual protocol.
Under our model and the choice of the step function $f_{\rm suc,0}(x)=\Theta(x-x_{\rm th})$ in Sec.~\ref{section-simulation}, it can be written as
\begin{equation}
 \begin{split}
P^{\pm} &\coloneqq \bra{\pm (-1)^a}  \mathcal{F}(\rho^a_{\rm model}) \ket{\pm (-1)^a}\\
&= \frac{1}{2}{\rm erfc} \left[(x_{\mathrm{th}}\mp\sqrt{\eta\mu})\sqrt{\frac{2}{2+\xi}}      \right],	
 \end{split}
\end{equation}
where
\begin{align}
{\rm erfc}(x)\coloneqq \frac{2}{\sqrt{\pi}}\int_{x}^{\infty}dt\: e^{-t^2}.
\end{align}
With $P^{\pm}$, we have
\begin{align}
   \mathbb{E}\left[   \hat N^{\rm suc}  \right] = Np_{\rm sig}\left(  P^{+}+P^{-}   \right),
\end{align}
which was used as the value of $\hat{N}^{\rm suc}$ in the simulation of the key rate.

For the cost $H_{\rm EC}$ of the error correction, we assume that the efficiency of the error correction is $1.1$.
It means that $H_{\rm EC}$ can be given by
\begin{align}
H_{\rm EC} &= 1.1\times \hat N^{\rm suc}h(e_{\rm bit}), 
\end{align}
with
\begin{align}
e_{\rm bit} &= \frac{P^{-}}{P^{+}+P^{-}}.
\end{align}
\newline

\section*{DATA AVAILABILITY}
Data sharing not applicable to the article as no datasets were generated or analyzed during the current study.
\section*{CODE AVAILABILITY}
Computer codes to calculate the key rates are available from the corresponding author upon reasonable request.
\section*{ACKNOWLEDGMENT}   
This work was supported by the Ministry of Internal Affairs and Communications (MIC) under the initiative Research and Development for Construction of a Global Quantum Cryptography Network (grant number JPMI00316);
Cross-ministerial Strategic Innovation Promotion Program (SIP) (Council for Science, Technology and Innovation (CSTI)); 
CREST(Japan Science and Technology Agency) JPMJCR1671;
JSPS  Grants-in-Aid  for  Scientific  Research No.~JP22K13977;
JSPS KAKENHI Grant Number JP18K13469.
\section*{AUTHOR CONTRIBUTIONS}
S.Y. and T.M. wrote the code for computing key rates and performed the calculations. Y.K. and T.S. contributed to the interpretation of the results. M.K. supervised the study. S.Y. drafted the original manuscript. All authors contributed to revising the manuscript.
\section*{COMPETING INTERESTS}
The authors declare no competing financial or non-financial interests.


\bibliographystyle{naturemag}
\bibliography{QKD}
\end{document}